\documentclass[12pt,preprint]{aastex}

\newcommand{\target}{2S\,0921--630/V395~Car}
\newcommand{\kms}{$\rm km\,s^{-1}$}
\newcommand{\Msun}{$\rm M_{\odot}$}
\newcommand{\Rsun}{$\rm R_{\odot}$}
\newcommand{\Lsun}{$\rm L_{\odot}$}

\slugcomment{}

\shorttitle{The massive neutron star or low-mass black hole in \target.}
\shortauthors{Shahbaz et al.}

\begin{document}

\title{The massive neutron star or low-mass black hole in \target.}

\author{T. Shahbaz\altaffilmark{1},
        J. Casares\altaffilmark{1},
        C. Watson\altaffilmark{2},
        P. A. Charles\altaffilmark{3},
        R. I. Hynes\altaffilmark{4},        
        S. C. Shih\altaffilmark{3},
        D. Steeghs\altaffilmark{5}
}

\altaffiltext{1}{Instituto de Astrof\'\i{}sica de Canarias, 38200 La Laguna,
Tenerife, Spain. tsh, jcv@iac.es}
\altaffiltext{2}{Department of Physics and Astronomy, University of Sheffield,
Sheffield, S3 7RH, England. c.watson@sheffeild.ac.uk}
\altaffiltext{3}{School of Physics and Astronomy,
The University of Southampton, Southampton, SO17 1BJ, UK. 
pac, icshih@soton.ac.uk }
\altaffiltext{4}{McDonald Observatory and Department of Astronomy,
The University of Texas at Austin, 1 University Station C1400, Austin, 
Texas 78712, USA. rih@astro.as.utexas.edu}
\altaffiltext{5}{Harvard-Smithsonian Center for Astrophysics,
60 Garden Street, MS-67, Cambridge, MA 02138, USA. steeghs@cfa.harvard.edu }
	
\begin{abstract}

We report on optical spectroscopy of the eclipsing Halo LMXB \target,
that reveals the absorption line radial velocity curve of the K0{\sc
iii} secondary star with a semi-amplitude $K_{\rm 2}$=92.89$\pm$ 3.84~\kms~, a
systemic velocity $\gamma$=34.9$\pm$3.3~\kms and an orbital period
$P_{\rm orb}$ of 9.0035$\pm$0.0029 day (1-$\sigma$).  Given the quality of 
the data we find no evidence for the effects of X-ray irradiation.  
Using the previously determined
rotational broadening of the mass donor, and applying conservative
limits on the orbital inclination, we constrain the
compact object mass to be 2.0--4.3~\Msun (1-$\sigma$), ruling out a
canonical neutron star at the 99\% level.  Since the nature of the
compact object is unclear, this mass range implies that the compact
object is either a low-mass black hole with a mass slightly higher
than the maximum neutron star mass (2.9~\Msun) or a massive neutron
star.  If the compact object in \target\ is a black hole, it confirms
the prediction of the existence of low-mass black holes, while if the
object is a massive neutron star its high mass severely constrains the
equation of state of nuclear matter.

\end{abstract}


\section{Introduction}

A knowledge of the neutron star mass distribution provides a
fundamental test of theories of the equation of state of nuclear
matter, the applicability of General Relativity as the correct theory
of gravity and significant information on the evolutionary history of
the progenitor stars. To date, only studies of millisecond radio
pulsars have provided accurate mass determinations of neutron stars,
reflecting their mass at formation; 1.35$\pm$0.04~\Msun\
\citep{Ker01}.  Although dynamical neutron star masses can also be
obtained from accreting X-ray pulsars, there are large
uncertainties. In high-mass X-ray binaries (HMXBs) there are
uncertainties due to non-Keplerian perturbations in the radial
velocity curves caused by effects such as stellar wind contamination
and X-ray heating.  In low-mass X-ray binaries (LMXBs), the situation
is worse because the intense X-ray irradiation usually
suppresses the light from the donor \citep{Cha03}.  It is only in a
few exceptional cases where the companion is evolved (and hence more
luminous) or during X-ray off-states, when dynamical information can
be extracted about the nature of the compact object.

The SAS-3 X-ray source, 2S~0921-630, was identified with a
$\sim$16$^m$ blue star \citep{Li78}, V395~Car, whose optical spectrum
is dominated by HeII\,4686\AA\ and Balmer emission \citep{Bra83},
characteristics of LMXBs, with occasional superimposed late-type
stellar absorption features visible \citep{Tho79}.  EXOSAT
observations showed a broad ($>$1~day), shallow X-ray eclipse, during
which the spectrum softened \citep{Mas87}, allowing an estimate of the
size of the accretion-disc corona (ADC).  This was comparable to the
eclipsing companion star, thus accounting for the partial X-ray
eclipse, and with an inclination of $i$=70$^\circ$--90$^\circ$.  
%The variability prior to the broad X-ray eclipse is reminiscent of
%2A\,1822-371 \citep{Whi85}, where obscuration of the ADC by structure
%on the disc rim is responsible for the shape of the lightcurve.  
Only a handful of ADC sources are known in which the compact object
is permanently obscured from our line-of-sight by the accretion disc,
and requires the observed X-rays to be scattered in a hot corona
\citep{Whi95}.  By implication, the intrinsic X-ray luminosity is 
{\it much} higher.

Limited optical photometry and emission line spectroscopy
\citep{Cow82,Bra83} shows that \target\ has a  long orbital period
of 9.02\,day and deep ($\sim$1 mag) dips where
the $(B-V)$ colour reddens by up to 0.4 mags, which has been
interpreted as the eclipse of the disc by the late-type
secondary \citep{Che82}.  High-resolution optical spectroscopy
revealed the donor star to be of spectral type  K0{\sc iii} with a
rotational velocity of $v\sin\,i$=65$\pm$9\,\kms, contributing
$\sim$25\%  to the observed flux at 6500\AA\ \citep{Sha99}. \target\
is one of  those rare LMXBs in which the secondary is visible despite
the presence of a luminous  disc.  
%Kinematic properties imply that
%it is located in the Halo at a distance $\sim$10\,kpc \citep{Cow82}.

There has been no detection of type {\sc i} X-ray bursts, so the
nature of the compact object is unclear. In this letter we present the
results of an intensive campaign to measure the dynamical mass of the
compact object in \target. We determine the secondary star's radial
velocity curve, investigate the possible effects of X-ray irradiation
and obtain firm constraints on the compact object mass.

\section{Observations and Data Reduction}

Time resolved spectroscopic observations of \target\ were obtained on
the  1.9-m telescope at the South African Astronomical Observatory
(SAAO) during 2003 March 30 to April 18.
Grating \# 5 (1200 lines/mm) was used centered at 4900\AA\ and
covering a wavelength range of 4529--5377\AA with a dispersion of 
0.49\,\AA/pixel$^{-1}$. 
A slit width of 2.0\,\arcsec gave a spectral  resolution of
1.1~\AA\ (=64~\kms\ at 5200\AA)  as measured from the  Cu-Arc arc
lines.    In general the conditions were good, with the seeing varying
between  1\,\arcsec--3\,\arcsec. A total of 83 spectra of \target\
were obtained with an exposure time of 1800\,s. Each spectrum was
bracketed by an observation of the internal Cu-Ar arc lamp at the
position of the star.  Template field stars of a variety of spectral
types were also observed.  

The images were flat-fielded using observations of a  tungsten calibration
lamp,  and a constant bias level was subtracted from all  the
images. Extraction of the 1-D spectra from the images were performed
using optimal extraction 
\citep{Hor86a} and calibration of the wavelength scale was achieved
using 4th order polynomial fits to the position measured arc-line
positions, which gave an rms scatter of $\sim$0.02~\AA.

In addition, 34 spectra were also obtained at the Very
Large Telescope (VLT; Paranal, Chile), New Technology Telescope 
(NTT; La Silla, Chile) the Anglo Australian Telescope (AAT; Siding Springs,
Australia)  and at the Magellan telescope (Las Campanas, Chile)
as part of a long-term project on X-ray binaries. 
We used the FORS2 Spectrograph attached to the 8.2m Yepun Telescope (VLT) 
on the nights of 2003 May 18 and June 22-23. 
A total of 9 spectra with exposure times in the range 300 to 900\,s were obtained.
The R1400V holographic grating was used in combination with a
0.7\,\arcsec slit resulting in a wavelength coverage of 4514--5815\AA\ and a
resolution of 70\,\kms. Instrumental flexure was monitored by
cross-correlating the sky spectra and was found to be very small, always 
within 6\,\kms.
The Magellan spectra were obtained on 2003 Dec 12--16 using 
the Baade telescope at Las Campanas using the IMACS imaging spectrograph in 
long-camera mode.
A total of 11 spectra with an exposure time of 300s were obtained.
The 600 grating together with a 0.7\,\arcsec long slit provided spectra
covering 3720--6830\AA\ with a dispersion of 0.75\,\AA/pixel$^{-1}$.
The AAT spectra were obtained using the RGO spectrograph on the nights of  2002 June 6 to 11.
A total of 10 spectra with an exposure time of 1800\,s were obtained.
The R1200B grating centered at 4350\AA\ together with a 1.0\,\arcsec slit provided spectra
covering 3500-5250\AA\ with a resolution of 70\,\kms.
The total of 4 NTT spectra were obtained with EMMI and grating \# 6  on the 
nights of 2002 June 8 to 10 exposure times of 600 to 900\,s.
The spectra cover the spectral range 4400--5150\AA\ with a resolution of 75\,\kms.
The AAT and NTT data reduction details are given in \citet{Cas03}.

\section{The radial velocity curve and spectrum}
\label{RadVel}

\begin{figure*}
\includegraphics[angle=90,scale=.80]{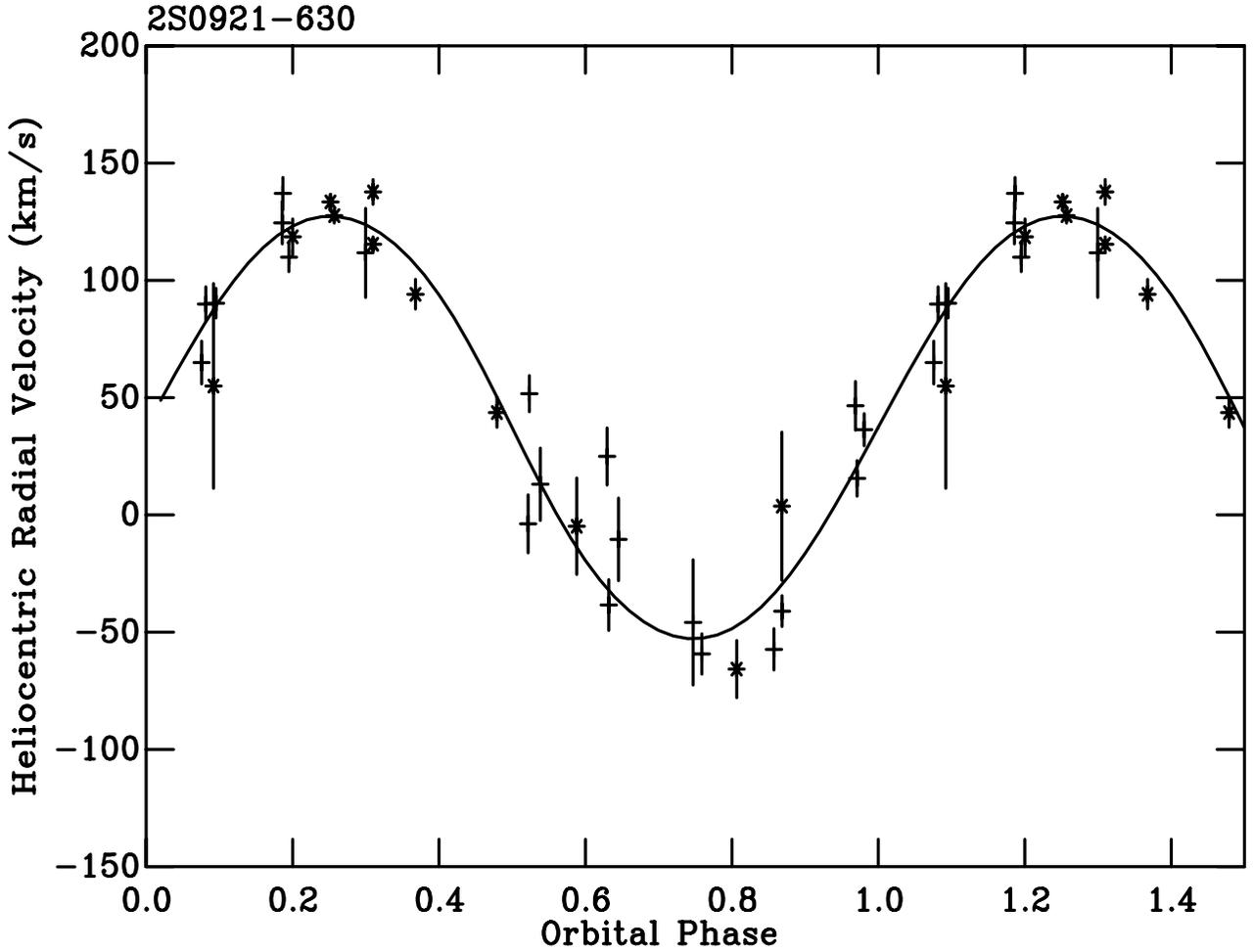}
\caption{The heliocentric radial velocity curve of the secondary star. The
crosses show the SAAO data and the stars the VLT, NTT, AAT and Magellan data.
The solid line shows a sinusoidal fit to the data. The data have been
folded on the orbital ephemeris and  1.5 orbital cycles are shown for clarity.
\label{RVCURVE}}
\end{figure*}

To increase the signal-to-noise of the spectra prior to
cross-correlation,  the individual spectra were variance
averaged into nighly means or groups of  $\sim$3\,hrs, depending on 
the quality of the individual spectra.  A total of 31
absorption line radial velocities were measured  by cross-correlation
\citep{Ton79} with a template star.  Prior to cross-correlation, the
spectra were interpolated onto a constant  velocity  scale (32
\kms\,pixel$^{-1}$) and  normalized by fitting a spline function to
the continuum.  The region 5100--5300\AA\ common to all the spectra,
primarily  containing
the Mg{\sc i} (5167.3\AA, 5172.7\AA\ and 5183.6\AA) absorption   blend
was used in the analysis.  The template star's  radial velocity,
determined using the position of the Fe{\sc i} 4957.597\AA\ absorption
line, was then added to the radial velocities of \target. The
resulting radial velocity data (see Figure\,\ref{RVCURVE}) were then
fitted with a circular orbit  (sinusoidal function) using the method
of least-squares.  Using spectral type template stars in the range
G6--K4{\sc iii} we obtain  $K_{\rm 2}$ values in the range 92--99~\kms
(see Table~\ref{RVFIT}).  The secondary star's spectral type has
previously been determined from high-resolution spectroscopy to be
K0{\sc iii} \citep{Sha99}, and our work confirms this.  Fitting the
radial velocities with a K0{\sc iii} template star, we find a minimum
reduced $\chi^2$ of 4.3 with the best fit parameters
 $\gamma$ = 34.9$\pm$3.3 \kms\, 
 $K_{\rm 2}$ = 92.89$\pm$3.84 \kms\,
 $P_{\rm orb}$=9.0035$\pm$0.0029 day and
 $T_{\rm 0}$=HJD\,2453099.51$\pm$0.08, 
where  $P_{\rm orb}$ is the orbital period, $T_{\rm 0}$ is time at
phase 0.0  defined as inferior conjunction  of the secondary star,
$\gamma$ is the systemic velocity and $K_{\rm 2}$ is the radial
velocity semi-amplitude (1-$\sigma$ errors are quoted with the error
bars rescaled so that the  reduced $\chi^2$ of the fit is 1.)  
Our ephemeris is consistent with that determined by \citealt{Mas87}.
We also
computed a $\chi^2$ periodogram of the data to investigate the
significance of other periods. No significant peaks (at $>$99.99 \%
level) were present other than the one at 9.0035\,day.  Finally, to
check if the 9.0035\,day period is affected by an alias, we  computed
the window function; no significant peaks were present at
9.0035\,day.  Fitting the radial velocity data with an eccentric
orbit did not yield a better fit; the eccentric fit is only
significant at the 38\% level.

Using a distance of  $\sim$10\,kpc  \cite{Cow82} \target\ lies 1.6\,kpc 
below the galactic plane and so belongs to the Halo population.   
Note the systemic velocity is not consistent with the radial velocity due to 
Galactic differential rotation,  $\sim$60~\kms\ at a distance 
of $\sim$10\,kpc  \citep{Deh98}. 
However, what is interesting is that the magnitude of the systemic velocity  
of \target\ seems to be about a factor
of $\sim$4 lower compared  to other  neutron star Halo systems 
(\citealt{Cas98,Cas02,Tor02}).

\begin{figure*}
\includegraphics[angle=90,scale=.80]{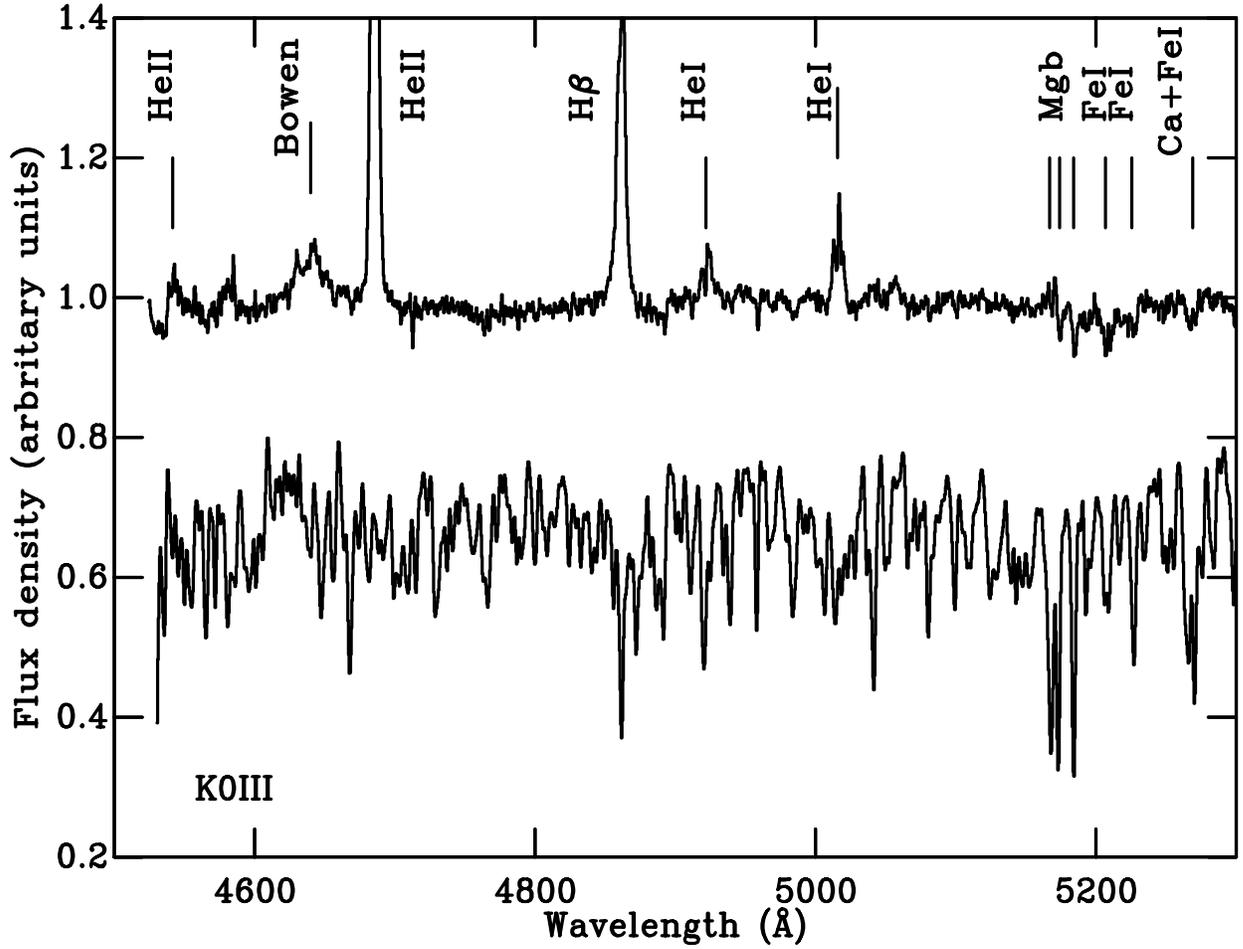}
\caption{
Top: the variance-weighted Doppler shifted average spectrum of \target\ 
in the rest frame of the secondary star. 
The most noticeable features are the H and He emission lines arising 
from the accretion disc.
Bottom: the K0{\sc iii}
spectral type template star HD82565. The spectra have been normalized and
shifted vertically for clarity.
\label{SPECTRUM}}
\end{figure*}

In Figure\,\ref{SPECTRUM} we show the variance-weighted Doppler
averaged spectrum of V395~Car. The strongest features are the H, He
emission lines, which appear single peaked, even at the resolution of
these data.  Weak absorption features of Mg {\sc i} 5175\AA\ and the
Ca/Fe blend at 5269\AA\ allow us to determine the fraction of light
from the secondary star, by optimally subtracting a scaled K0{\sc iii}
template spectrum from V395~Car \citep{Mar94}.  As the rotational
broadening of the donor (65$\pm$9~\kms; \citealt{Sha99}) is similar to
our spectral resolution (64~\kms), we do not broaden the template
before subtraction. We find that the secondary star contributes $\sim
10$\% to the observed flux at 5200\AA\ (see Table~\ref{RVFIT}).

\section{Irradiation of the secondary star - the mass of the compact object}
\label{Irrad}

\begin{figure*}
\includegraphics[scale=.35]{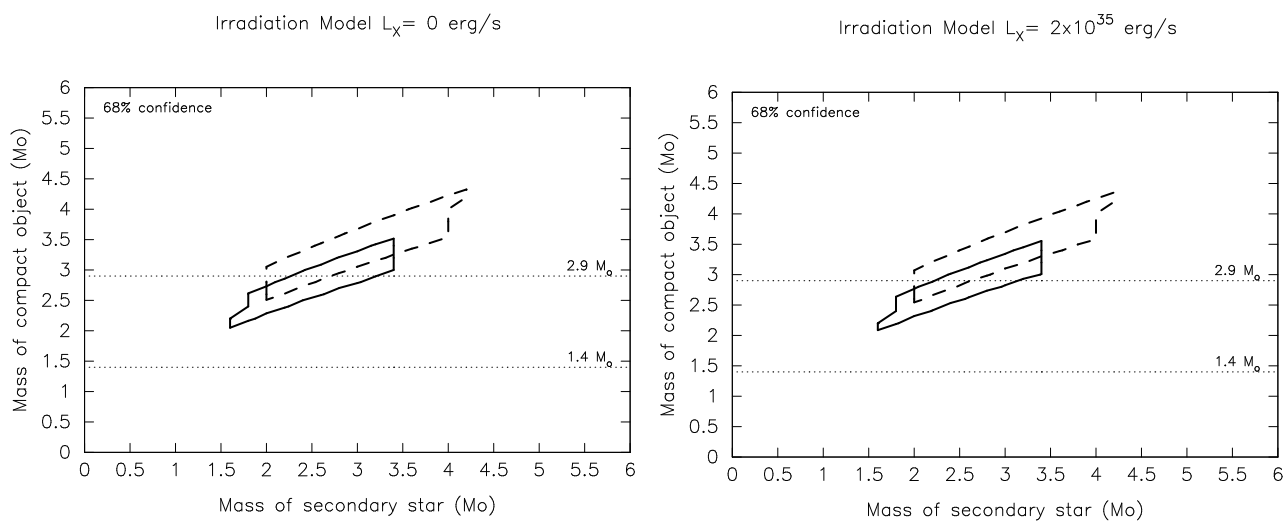}
\caption{
The effects of X-ray heating on the secondary star's radial velocity curve and
hence the allowed binary component masses. The left and right plots show the
case without and with X-ray heating respectively.
The solid and dashed
contours show the 68\% confidence level obtained by fitting the radial velocity
curve with the X-ray binary model using $i$=90$^\circ$ and 70$^\circ$
respectively (see section\,\ref{Irrad})
The dotted horizontal lines mark 1.4~\Msun\ and 2.9~\Msun, and 
represent the canonical and maximum neutron star mass respectively.
\label{MASSES}}
\end{figure*}

The observed X-rays are reflected into our line of sight by
scattering.  Although the observed X-ray luminosity is low  
$L_{\rm X}=2\times 10^{35}$\,erg\,s$^{-1}$ \citep{Kal03}, it is still
interesting to explore the possible effects that X-rays may have on
the secondary star's atmosphere and hence radial velocity.
We investigate these effects using the X-ray binary model described by
\citet{Phi99,Sha00,Sha03}.  We first phase-folded the
radial velocities using the orbital ephemeris derived in
section\,\ref{RadVel} and  binned into 31 orbital phase bins.  
We then fit the resulting radial velocity curve with the X-ray binary model.
We considered two extreme cases for the X-ray irradiation: case (a)
where the X-rays produced near the compact object irradiate the
whole inner face of the secondary star  and case (b) where the X-ray's
do not affect the secondary star.  Note that invoking a flared
accretion disc will have intermediate solutions. The
variable model parameters are the compact object mass  $M_{\rm 1}$,
secondary star mass $M_{\rm 2}$, X-ray luminosity $L_{\rm X}$  and the
inclination $i$.   To determine which elements on the
star contribute to the absorption line  radial velocity we use the
factor  $f_{\rm X}$, which is the  fraction of the external radiation
flux that exceeds the unperturbed flux.  
For a given $i$ and
combination of   $M_{\rm 1}$ and $M_{\rm 2}$ we fit the data and use
the  constraints imposed by the measured value for $v\sin\,i$ 
(64$\pm$9\,\kms; \citealt{Sha99}) to
produce plots in the  $M_{\rm 1}$--$M_{\rm 2}$
plane (Figure\,\ref{MASSES})

Given the evidence for optical and X-ray partial eclipse, $i$ must lie
in the range 70$^\circ$--90$^\circ$.  Therefore we consider cases (a)
and (b) for $i$=70$^\circ$ and more importantly $i$=90$^\circ$,
because it gives a firm lower limit to the compact object's mass. For
case (a) we fix $L_{\rm X}=2\times 10^{35}$\,erg\,s$^{-1}$ and find
that the best fit ($\chi^2$ = 122) is obtained when all the elements
on the inner face contribute to the radial velocity, i.e. $f_{\rm X}>$
60\%, i.e. X-ray heating is not preferred.  Note, however, that the
radial velocity curve will not be sinusoidal because there will still
be some irradiated elements that contribute to the radial velocity.
Even if we increase to $L_{\rm X}=10^{39}$\,erg\,s$^{-1}$ as could be 
the case  for an LMXB, the fits are significantly worse ($\chi^2$ = 190) 
and the best fit still occurs when all the inner face contributes to the 
radial velocity. For case (b) we set $L_{\rm X}=0$, implying no X-ray 
heating and find that it gives a fit with
similar quality ($\chi^2$ = 119).  We conclude that the best fit to
the radial velocity curve is the case where the X-rays do not have an
observable effect given the quality of the data.
For our range in $i$ our fits for case (a) and (b)
give 2.0 $< M_{\rm 1} <$ 4.3~\Msun\ (1-$\sigma$), which means we can
rule out a canonical 1.4\,\Msun\ neutron star at the 99\% level.  Our
solutions for the compact object mass suggests
either a massive neutron star or a low-mass black hole, since the
maximum mass allowed for a neutron star is 2.9~\Msun\ \citep{Kal96}.

Since the donor star fills its Roche lobe  and is synchronized
with the binary motion, we use the secondary star's $v\sin\,i$ 
and $K_{\rm 2}$ to determine the binary mass ratio 
$q$=($M_{\rm 2}/M_{\rm 1}$)=0.89$\pm$0.18 \citep{Hor86b}. Given the
measured masses and orbital period, we use Kepler's Third Law to
determine the semi-major axis 56.1$<a<$72.4\,\Rsun.
Eggleton's (1983) expression for the effective radius of the  Roche-lobe 
then determines the radius of the secondary 10.4$<R_{\rm 2}<$13.4\,\Rsun,
and the temperature inferred from the spectral type  \citep{Gra92}
and   Stefan-Boltzmann's law determines the luminosity  
52.5$<L_{\rm 2}<$87.1\,\Lsun (1-$\sigma$ ranges are quoted).

\section{Discussion}

\subsection{The evolutionary status of the system}
\label{Evol}

The position of the secondary star of \target\ in an HR diagram corresponds to a
normal star that has crossed the Hertzsprung gap and now lies on the
Hayashi line.  The evolution of the binary is dominated by the
evolution of the evolved secondary star. Since such a star no longer
burns hydrogen in its core, the mass transfer is early massive Case B.
For a binary with a secondary star near the onset of such mass
transfer, one requires $q<$1 throughout its history \citep{Kin99}.
The case for a 1.4~\Msun\ neutron star can be ruled out since it
requires $M_{\rm 2}<$1.4~\Msun, which is inconsistent with our
results; the position of a secondary on the HR diagram undergoing
early massive Case B mass transfer is close to that of a single star
of the same mass \citep{Kol98}. For a 3~\Msun\ compact object, $M_{\rm
2}<$3.0~\Msun, which is consistent with our observations.

It is interesting to note that the mass ratio we derive lies on the limit
of stable/unstabe mass transfer $q$=5/6 \citep{Kin99}. It is possible that \target\ 
is undergoing unstable mass transfer similar to the supersoft sources and other 
X-ray binaries  \citep{Kah97}.

\subsection{The formation of the compact object}

The distribution of measured neutron stars masses provides fundamental 
constraints on the equation of state of nuclear matter and the 
observational identification of a black hole which is based on the  maximum
gravitational mass allowed for a neutron star;  2.9~\Msun\ \citep{Kal96}. 
Studies of radio pulsars  show that neutron star masses are clustered in a
narrow range. However, there is
some evidence that neutron stars with a mass in excess of 1.4~\Msun\ do
exist. LMXBs are expected to contain massive  ($\sim$2~\Msun) neutron stars, 
resulting from the accretion of a considerable amount of material over
extended  (10$^8$~yrs) periods of time \citep{Zha97}.
However, while a neutron star mass of  1.78$\pm$0.23~\Msun\ for the LMXB 
Cyg\,X-2 is reported \citep{Oro99},  \citet{Tit02} have recently proposed a
mass of 1.44$\pm$0.06~\Msun\ based on the spectral and temporal properties of
the type {\sc I} X-ray bursts. The analysis of optical data for the X-ray
pulsar Vela\,X-1 suggests that the pulsar has a mass  of
1.87$^{+0.23}_{-0.17}$~\Msun\ \citep{Bar02}. Probably the strongest case for a
massive neutron star or low-mass black  hole is in the eclipsing HMXB
4U\,1700--37 which contains a  2.44$\pm$0.27\,\Msun\ compact object
\citep{Cla02}. Such masses already test soft 
nuclear equations of state \citep{Mil98a}.
We obtain a mass of 2.0--4.3~\Msun\ for the compact object in \target, which 
lies between the range of masses observed for neutron stars and black holes.
Models for nuclear equations of state which include the  effects of three
nucleon interactions and  realistic models of nuclear forces limits
the maximum mass of neutron stars to be below 2.5~\Msun \citep{Akm98}.
However, it should be noted that unconventional forms of matter such as
Q-stars do allow the existence of extremely massive neutron stars \citep{Mil98b}
Although our results do not allow us to distinguish between a massive neutron
star or a low mass black hole, the existence of a $>$2.5~\Msun neutron star 
would place strong constraints on high density nuclear matter. 

Theoretical predictions by \cite{Fry01} suggest that the death of a
60~\Msun\ star in a close binary system can produce anything from a
low-mass 1.2~\Msun\ neutron star to a 10~\Msun\ black hole, depending
on the wind mass-loss rate during the Wolf-Rayet phase.  As suggested
by \citet{Bro96} the formation of a low-mass black hole with a high 
space velocity is  possible via a two stage process involving 
the formation of a neutron star.
However, if there is symmetric mass ejection in the supernova, then a kick at
the formation of a black hole is not required to explain the varied
range in the observed space velocities of the black hole LMXBs
\citep{Nel99}.

The unusual mass for the compact object in \target\ and the fact that the 
nature of the compact object is not clear, conjours up many formation
scenarios. If it is a
neutron star, then its present mass  can be explained by either the
accumulation of matter (1--2~\Msun) by a  canonical 1.4~\Msun\ neutron star,
accreted at a high mass accretion rate over a long period of time, or the
direct collapse of a massive star forming a massive neutron star.  If the
compact object in \target\ is a black hole then its low mass can be explained
by the accretion induced collapse of a massive 2.9~\Msun\ neutron star or the
formation of a low-mass black hole \citep{Bro96}.  Given the fact that we do
not observe a peculiar systemic velocity comparable to other
neutron-star LMXBs, this suggests that we can rule out canonical neutron star
formation scenarios where a kick is produced during a type {\sc II} 
supernova. The most likely scenario for the formation of the compact object is
the direct formation of a massive neutron star or a low-mass black hole.

\acknowledgments

TS and JC acknowledge support from the Spanish Ministry of Science 
and Technology under the programme Ram\'{o}n y Cajal.
RIH is currently supported by NASA through Hubble Fellowship grant
\#HF-01150.01-A\ awarded by the STSci, which
is operated by AURA, for NASA, under contract NAS 5-26555.
DS acknowledges a Smithsonian Astrophysical Observatory Clay Fellowship.
The data reduction and analysis was performed using the {\sc pamela} 
and  {\sc molly} routines of K.\,Horne and T.\,R.\ Marsh.
This paper uses observations made at the South African Astronomical Observatory
(SAAO).

%Facilities: \facility{SAAO}, \facility{AAT}, \facility{NTT}, \facility{VLT},
%\facility{Magellan}

\clearpage

%% This example uses \plotone to include an EPS file scaled to
%% 80% of its natural size with \epsscale. Its caption
%% has been written to indicate that additional figure parts will be
%% available in the electronic journal.

%% Here we use \plottwo to present two versions of the same figure,
%% one in black and white for print the other in RGB color
%% for online presentation. Note that the caption indicates
%% that a color version of the figure will be available online.
%%

\begin{table}
\begin{center}
\caption{Fits to the radial velocity curve of \target\ (1-$\sigma$ errors are given).
\label{RVFIT}}
\begin{tabular}{lccccccc}
\\
\tableline\tableline
Template & SpT & $\gamma$ & $P_{\rm orb}$ & $T_{\rm 0}$ $*$& $K_{\rm 2}$ & $\chi^2_{\nu}$ & $f$ \\
         & & (\kms) & day) & & (\kms) & & \% \\
\tableline
HD122571 & G6III & 33.3 (3.8) & 9.0019(36) & 9.54(10)    &  95.1(4.4)  & 5.8  &  6.7 \\
HD63513  & G7III & 34.2 (3.6) & 9.0017(32) & 9.53(9)     &  94.2(3.7)  & 5.3  &  7.8 \\
HD40359  & G8III & 36.9 (3.1) & 9.0040(30) & 9.51(9)     &  92.2(3.7)  & 4.5  &  8.6 \\
HD71863  & G9III & 35.1 (3.3) & 9.0027(29) & 9.51(8)     &  93.9(3.8)  & 4.3  &  9.6 \\
HD82565  & K0III & 34.9 (3.3) & 9.0035(29) & 9.51(8)     &  92.9(3.8)  & 4.3  &  9.9 \\
HD39523  & K1III & 37.6 (3.2) & 9.0036(28) & 9.53(7)     &  94.3(3.7)  & 4.6  &  9.0 \\
HD61248  & K3III & 41.0 (3.5) & 9.0062(32) & 9.58(7)     &  96.5(4.1)  & 6.1  &  7.9 \\
HD40522  & K4III & 42.0 (3.7) & 9.0057(34) & 9.59(7)     &  98.6(4.3)  & 6.0  &  8.2 \\
\tableline
\end{tabular}
\end{center}
$*$HJD 2453099+
\end{table}

\end{document}